\documentstyle[12pt,a4,sw20lart]{article}

\input tcilatex
\QQQ{Language}{
American English
}

\begin{document}

\title{On the PPN 1+2 Body Problem\thanks{%
GRL Report 12/04/1997}}
\author{D. \c Selaru, I. Dobrescu \\
{\it Gravitational Research Laboratory, 21-25 Mendeleev str., }\\
{\it 70168 Bucharest, Romania}\\
e-mail: {\it dselaru@scou1.mct.ro, idobrescu@scou1.mct.ro}}
\maketitle

\begin{abstract}
A particular case of the three-body problem, in the PPN formalism, is
presented. The Hamiltonian function is obtained and the problem is reduced
to a perturbed two-body one.
\end{abstract}

\section{INTRODUCTION}

Several approaches were performed on the PPN two-body problem, (Soffel,
1989). In the PPN three-body problem the most studied case is the motion of
a test particle in the gravitational field of two massive bodies, i. e. the
restricted problem (Brumberg, 1972). This paper presents a different
situation: the motion of a binary system of two very small masses around a
massive body, when the motion of the massive body not is affected by the
presence of the couple. The physical situations modeled may be the motion of
a binary asteroid (a binary comet fragments) or of two artificial objects.
The special choice of masses (the small bodies have equal masses)
particularize the last case. One uses the PPN formalism in the first
approximation in the Hamiltonian formulation. In the {\em Jacobi}
coordinates, the three-body problem splits in two coupled perturbed two-body
problems: the motion of the barycenter of the binary system and the relative
motion of the two small bodies. The first is rather well approximated by the
motion of a test particle in the gravitational field of a massive body, e.
g. Damour and Deruelle (1985), Heimberger {\em et al}. (1990). An imposed
circular solution for this problem is substituted in the two-body relative
motion. Similar to the case of the restricted three-body problem, one passes
to a comoving coordinate system, where the $Ox$ axis coincides with the
direction given by the mass center of the binary system and the center of
the massive body. In this system the Hamiltonian becomes autonomous.
Conclusions are given to the possibility of a GRT experiment in this
framework.

We start with the usual {\em Lagrange} function $L_N$ of the N-particles
moving in a self-consistent field in the PPN formalism (see Soffel, 1989 for
notations). We obtain the {\em Hamilton} function for the three-body problem
by means of the {\em Legendre} transformations: 
\[
H_3=\sum_i^3<{\bf p_i,v_i}>-L_3\text{ with }{\bf p_i}=\frac{\partial L_3}{%
\partial {\bf v_i}}\Longrightarrow 
\]
\begin{equation}
H_3=\sum_i^3\left( \frac{p_i^2}{2m_i}-\frac{p_i^4}{8c^2m_i^3}\right) -\frac
12\sum_{i,j\neq i}^3\frac{Gm_im_j}{r_{ij}}-\sum_{i,j\neq i}^3\frac{G(2\gamma
+1)m_j}{2c^2m_ir_{ij}}p_i^2+  \label{1}
\end{equation}
\[
+\sum_{i,j\neq i}^3\frac G{4c^2}\left( \frac{4\gamma +3}{r_{ij}}<{\bf %
p_i,p_j>+}\frac 1{r_{ij}^3}<{\bf p_i,r_{ij}><p_j,r_{ij}>}\right) + 
\]
\[
+\frac{2\beta -1}{2c^2}\sum_{i,j\neq i}^3\sum_{k\neq i}^3\frac{G^2m_im_j}{%
r_{ij}}\frac{m_k}{r_{ik}}. 
\]
The Hamiltonian formalism is used in order to benefit of the {\em Lie-Deprit}
method in the study of the near integrable systems.

\section{THE\ PPN\ 1+2 BODY\ PROBLEM WITH\ EQUAL SMALL MASSES}

This section describes the Hamiltonian of a particular case of the
three-body problem: a close pair, $m_1=m_2=m,$ rotating around the common
center of mass and the third body, $m_3=M,$ far from these two, with $m<<M$.
We perform the simplectic transformations ({\em Jacobi} coordinates): 
\begin{equation}
{\bf r_1}=\frac{{\bf r}}2+{\bf R}+{\bf R^{\prime }},\;{\bf r_2}=-\frac{{\bf r%
}}2+{\bf R}+{\bf R^{\prime }},\;{\bf r_3}={\bf R^{\prime }},  \label{2}
\end{equation}
\[
{\bf p_1}={\bf p_r}+\frac{{\bf p_R}}2,\;{\bf p_1}=-{\bf p_r}+\frac{{\bf p_R}}%
2,\;{\bf p_3}=-{\bf p_R}+{\bf p_{R^{\prime }}.} 
\]
Due to that ${\bf R^{\prime }}$ is not present in the Hamiltonian function
(consequently, ${\bf p_{R^{\prime }}}$ is a constant of motion) and
considering ${\bf p_{R^{\prime }}}=0,$ we obtain: 
\[
H_{1+2}=H_{Newton}+H_{PPN}, 
\]
where 
\begin{equation}
H_{Newton}=\frac{p_r^2}m-\frac{Gm^2}r+\frac{p_R^2(M+2m)}{4mM}-\frac{2GmM}R-
\label{3}
\end{equation}
\[
-GmM\left( \frac 1{\left| \frac{{\bf r}}2-{\bf R}\right| }+\frac 1{\left| 
\frac{{\bf r}}2+{\bf R}\right| }-\frac 2R\right) 
\]
and 
\begin{equation}
H_{PPN}=-\frac 1{8c^2m^3}\left( 2p_r^4+\frac{p_R^4}8+p_R^2p_r^2+2<{\bf %
p_r,p_R}>^2\right) -\frac{p_R^4}{8c^2M^3}-  \label{4}
\end{equation}
\[
-\frac{G(2\gamma +1)}{2c^2r}(2p_r^2+\frac{p_R^2}2)-\frac{G(2\gamma +1)M}{%
2c^2m}\left[ \left( p_r^2+\frac{p_R^2}4+<{\bf p_r,p_R}>\right) \frac
1{\left| \frac{{\bf r}}2+{\bf R}\right| }+\right. 
\]
\[
\left. +\left( p_r^2+\frac{p_R^2}4-<{\bf p_r,p_R}>\right) \frac 1{\left| 
\frac{{\bf r}}2-{\bf R}\right| }\right] -\frac{G(2\gamma +1)mp_R^2}{2c^2M}%
\left( \frac 1{\left| \frac{{\bf r}}2-{\bf R}\right| }+\frac 1{\left| \frac{%
{\bf r}}2+{\bf R}\right| }\right) + 
\]
\[
+\frac G{2c^2}\left[ \frac{4\gamma +3}r\left( -p_r^2+\frac{p_R^2}4\right)
+\frac 1{r^3}\left( <\frac{{\bf p_R}}2,{\bf r}>^2-<{\bf p_r,r}>^2\right)
\right] + 
\]
\[
+\frac G{2c^2}\left[ \frac{4\gamma +3}{\left| \frac{{\bf r}}2+{\bf R}\right| 
}\left( -<{\bf p_r,p_R}>-\frac{p_R^2}2\right) +\frac{<\frac{{\bf p_R}}2+{\bf %
p_r},\frac{{\bf r}}2+{\bf R}><-{\bf p_R},\frac{{\bf r}}2+{\bf R}>}{\left| 
\frac{{\bf r}}2+{\bf R}\right| ^3}\right] + 
\]
\[
+\frac G{2c^2}\left[ \frac{4\gamma +3}{\left| \frac{{\bf r}}2-{\bf R}\right| 
}\left( <{\bf p_r,p_R}>-\frac{p_R^2}2\right) +\frac{<\frac{{\bf p_R}}2-{\bf %
p_r},\frac{{\bf r}}2-{\bf R}><-{\bf p_R},\frac{{\bf r}}2-{\bf R}>}{\left| 
\frac{{\bf r}}2-{\bf R}\right| ^3}\right] + 
\]
\[
+\frac{(2\beta -1)G^2m}{c^2}\left[ \frac{m^2}{r^2}+\frac{mM}r\left( \frac
1{\left| \frac{{\bf r}}2-{\bf R}\right| }+\frac 1{\left| \frac{{\bf r}}2+%
{\bf R}\right| }\right) +\right. 
\]
\[
\left. +\frac{M^2+mM}2\left( \frac 1{\left| \frac{{\bf r}}2-{\bf R}\right|
}+\frac 1{\left| \frac{{\bf r}}2+{\bf R}\right| }\right) ^2-\frac{M^2}{%
\left| \frac{{\bf r}}2-{\bf R}\right| \left| \frac{{\bf r}}2+{\bf R}\right| }%
\right] . 
\]
We have 
\begin{equation}
\frac 1{\left| \frac{{\bf r}}2-{\bf R}\right| }=\sum_{n=0}^\infty \frac{r^n}{%
2^nR^{n+1}}P_n\left( \frac{<{\bf r,R}>}{rR}\right) ,\;\frac 1{\left| \frac{%
{\bf r}}2+{\bf R}\right| }=\sum_{n=0}^\infty \frac{r^n}{2^nR^{n+1}}%
(-1)^nP_n\left( \frac{<{\bf r,R}>}{rR}\right) ,  \label{5}
\end{equation}
where$\;P_n$ is the n-{\em th} usual {\em Legendre} polynomial.

\section{THE DECOUPLING OF THE TWO PROBLEMS}

A physical realization for this problem is the motion of two bodies with
masses of 10$^3$ kg each, with 1 m relative semi-major axis (considered
unperturbed) and situated at the distance of 20 mil. km. from the {\em Sun}.
The numerical evaluations for the perturbative terms from the Eqs. (\ref{3})
and (\ref{4}) will be estimated in this case. For the {\em Hamilton}
function given by Eq. (\ref{4}) the Newtonian orders of perturbations for
the relative problem $(m_1,\;m_2)$ are: 
\begin{equation}
\lambda _1=\frac M{2m}\left( \frac rR\right) ^3=O\left( 10^{-4}\right)
\label{6}
\end{equation}
and for the motion of the barycenter relative to the massive body: 
\begin{equation}
\lambda _2=\frac 14\left( \frac rR\right) ^2=O\left( 10^{-21}\right) .
\label{7}
\end{equation}
In this case, the motion of the barycenter of the couple might be considered
unperturbed in a good approximation, $\lambda _2=O(\lambda _1^5)$. The
problem is reduced to the study of the relative motion in the binary system.
The motion of the massive body is not affected by the presence of the
couple; this type of problem being considered as ''1+2 body problem''. By
retaining the significant terms, the problem splits in two perturbed
two-body problems. The motion of the barycenter is described by the
following Hamiltonian: 
\begin{equation}
H_{cm}=\frac{p_R^2(M+2m)}{4mM}-\frac{2GmM}R-\frac{p_R^4}{64c^2m^3}-\frac{%
G(2\gamma +1)M}{4c^2m}\frac{p_R^2}R+  \label{8}
\end{equation}
\[
+\frac{(2\beta -1)G^2mM^2}{2c^2R^2}+O\left( 10^{-20}\frac{p_R^2}{4m}\right)
. 
\]
The perturbing terms maintained (of the order of $O(10^{-8}p_R^2/4m)$) from
Eq. (\ref{8}) are given by the PPN model of the problem, which decouples
from the relative problem in the binary system in the considered
approximation. The resulting Hamiltonian is similar to the one for the case
of the motion of a test particle in the gravitational field of a massive
body. This problem admits a circular solution (Krefetz, 1967). In polar
coordinates (the reference plane is that of the binary system barycenter
motion) 
\[
R=\sqrt{X^2+Y^2},\;P_R,\;\Phi =\arctan \frac YX,\;P_\Phi , 
\]
(the canonic change of coordinates is obvious) this solution has the form: 
\begin{equation}
R=R_0^{\prime },\;P_R=0,\;P_\Phi =P_{\Phi 0}^{\prime },\;\Phi =n^{\prime
}t+\Phi _0.  \label{9}
\end{equation}
One denotes with ' the initial conditions from the Eq. (\ref{9}), in order
to distinguish between them and those of the Newtonian problem, (without
losing the generality, one may consider $\Phi _0^{\prime }=0$). The relative
motion in the binary system is described by the following Hamilton function: 
\[
H_{bin}=\frac{p_r^2}m-\frac{Gm^2}r-\frac{GmMr^2}{2R^3}P_2\left( \frac{<{\bf %
r,R}>}{rR}\right) - 
\]
\begin{equation}
-\frac 1{8c^2m^3}\left( p_R^2p_r^2+2<{\bf p_r,p_R}>^2\right) +\frac{Gp_R^2}{%
8c^2r}-\frac{G(2\gamma +1)Mp_r^2}{c^2mR}-  \label{10}
\end{equation}
\[
-\frac{G(2\gamma +1)Mr}{c^2mR^2}<{\bf p_r,p_R}>+\frac{G<{\bf r,p_R}>^2}{%
8c^2r^3}+\frac{2(2\beta -1)G^2m^2M}{c^2Rr}+O\left( 10^{-12}\frac{p_r^2}%
m\right) . 
\]
Introducing Eq. (\ref{9}) in Eq. (\ref{10}) one obtains: 
\[
H_{bin}=\frac{p_r^2}m\left( 1-\frac{P_{\Phi 0}^{\prime 2}}{%
8c^2m^2R_0^{\prime 2}}-\frac{G(2\gamma +1)M}{c^2R_0^{\prime }}\right) -\frac{%
Gm^2}r\left( 1-\frac{P_{\Phi 0}^{\prime 2}}{8c^2m^2R_0^{\prime 2}}-\frac{%
2(2\beta -1)GM}{c^2R_0^{\prime }}\right) - 
\]
\begin{equation}
-\frac{GmMr^2}{2R_0^{\prime 3}}P_2\left( \frac{x\cos (n^{\prime }t)+y\sin
(n^{\prime }t)}r\right) -\frac{P_{\Phi 0}^{\prime 2}}{4c^2m^3R_0^{\prime 2}}%
\left( -p_x\sin (n^{\prime }t)+p_y\cos (n^{\prime }t)\right) ^2-  \label{11}
\end{equation}
\[
-\frac{G(2\gamma +1)P_{\Phi 0}^{\prime }Mr}{c^2mR_0^{\prime 3}}\left(
-p_x\sin (n^{\prime }t)+p_y\cos (n^{\prime }t)\right) + 
\]
\[
+\frac{GP_{\Phi 0}^{\prime 2}}{8c^2r^3R_0^{\prime 2}}\left( -x\sin
(n^{\prime }t)+y\cos (n^{\prime }t)\right) ^2+O\left( 10^{-12}\frac{p_r^2}%
m\right) . 
\]
The relativistic terms retained are of the order of $O(10^{-8}p_r^2/m)$.
Passing to the comoving coordinate system (which rotates such that the axis $%
Ox$ joints the massive body with the barycenter of the binary system), we
obtain the autonomous Hamiltonian: 
\[
H_{bin}=\frac{p_r^2}m\left( 1-\frac{P_{\Phi 0}^{\prime 2}}{%
8c^2m^2R_0^{\prime 2}}-\frac{G(2\gamma +1)M}{c^2R_0^{\prime }}\right) -\frac{%
Gm^2}r\left( 1-\frac{P_{\Phi 0}^{\prime 2}}{8c^2m^2R_0^{\prime 2}}-\frac{%
2(2\beta -1)GM}{c^2R_0^{\prime }}\right) + 
\]
\begin{equation}
+n^{\prime }\left( xp_y-yp_x\right) -\frac{GmM}{4R_0^{\prime 3}}\left(
3x^2-r^2\right) -\frac{P_{\Phi 0}^{\prime 2}}{4c^2m^3R_0^{\prime 2}}p_y^2+
\label{12}
\end{equation}
\[
-\frac{G(2\gamma +1)P_{\Phi 0}^{\prime }M}{c^2mR_0^{\prime 3}}rp_y+\frac{%
GP_{\Phi 0}^{\prime 2}}{8c^2R_0^{\prime 2}}\frac{y^2}{r^3}+O\left( 10^{-12}%
\frac{p_r^2}m\right) . 
\]
The influence of the PPN model of the 1+2 problem is:

a) a direct one due to the presence of the last three terms in Eq. (\ref{12}%
),

b) an indirect one, by the modification of the coefficients which appear in
the other terms, also present in the Newtonian case.

\section{CONCLUSIONS}

The main purpose of this paper is to present a theoretical problem which may
constitute the premise of a GRT experiment. The prelimined development of
Solar System space missions could create the technological premises for
setting up such a 1+2{\em \ }body system, formed of equal masses, on a
precise given orbit, with the purpose to ''detect'' the influence of some
PPN parameters in order to evaluate the GRT predictions. To become real,
this model should take into account facts as: the presence of other bodies,
the shape of the massive body, a more realistic orbit of the barycenter than
the circular one, the radiation pressure, the influence of the
electromagnetic field, etc.

The study of this type of problem has some advantages:

- the possibility of choosing the appropriate initial conditions: the masses
rapport, the semi-major axis of the binary system mass center motion
relative to the massive body, the orbital elements of the relative motion of
the two-body system related to the orbital plane of the mass center motion;

- the ''faster evolution of the time'' in the sense that one needs a shorter
interval of time for performing the experiment than in the case of the study
of a natural body motion;

- the possibility of realizing more precise measurements for the obtained
effects, due to the fact that the bodies in the binary system are very close
and due to the high precision given by the present measuring devices which
may be set up on-board.

One notices here that, for the chosen initial data, in order to obtain the
relativistic correction for the indirect effects, it is necessary a first
order theory with the small parameter $\lambda _1$. In order to obtain the
direct effects a second order theory should be performed by means of
standard procedures.

\end{document}